\documentclass[
	aip, 
	prb,
	superscriptaddress, 
	reprint,
	nolinenumbers,
	longbibliography,
	amsmath,amssymb,floatfix.xcolor=table,
]{revtex4-2}

\usepackage{graphicx}
\usepackage{siunitx}

\usepackage[utf8]{inputenc}
\usepackage[T1]{fontenc}

\renewcommand*{\epsilon}{\varepsilon}

\DeclareTextCommand{\permil}{T1}{\%\char24}

\newcommand{\ccw}{\ensuremath{\circlearrowleft}} 
\newcommand{\cw}{\ensuremath{\circlearrowright}} 

\usepackage{xcolor}

\newcommand\hidden[1]{{}}

\newcommand{\JNN}{\ensuremath{J_\mathrm{NN}^\text{dip}}}
\newcommand{\JNNN}{\ensuremath{J_\mathrm{NNN}^\text{dip}}}
\newcommand{\kB}{\ensuremath{k_\mathrm{B} } }
\newcommand{\TN}{\ensuremath{T_\mathrm{N}}}

\makeatletter
\def\maketitle{
	\@author@finish
	\title@column\titleblock@produce
	\suppressfloats[t]}
\makeatother
\newcommand{\beginsupplement}{
	\setcounter{page}{1}
	\setcounter{section}{0}
	\renewcommand{\thesection}{S\arabic{section}}%
	\setcounter{table}{0}
	\renewcommand{\thetable}{S\arabic{table}}%
	\setcounter{figure}{0}
	\renewcommand{\thefigure}{S\arabic{figure}}%
	\setcounter{equation}{0}
	\renewcommand{\theequation}{S\arabic{equation}}%
	}

\newcommand{\papertitle}{Relaxation pathways and emergence of domains in square artificial spin ice}

\newcommand{\nanogune}{CIC nanoGUNE BRTA, 20018 Donostia - San Sebasti\'{a}n, Spain}
\newcommand{\ikerbasque}{IKERBASQUE, Basque Foundation for Science, E-48009 Bilbao, Spain}
\newcommand{\pancaldi}{Elettra-Sincrotrone Trieste S.C.p.A., 34149 Basovizza, Trieste, Italy}
\newcommand{\lboro}{Department of Physics, School of Science, Loughborough University, LE11 3TU Loughborough, United Kingdom}
\newcommand{\pg}{Department of Physics and Astronomy, University of British Columbia, Vancouver, BC V6T 1Z1, Canada}

\begin{document}

\title{\papertitle}

\author{Matteo Menniti}
\affiliation{\nanogune}

\author{Na\"emi Leo}
\affiliation{\nanogune}
\affiliation{\lboro}
\email{n.leo@lboro.ac.uk}

\author{Pedro Villalba-González}
\affiliation{\nanogune}
\affiliation{\pg}

\author{Matteo Pancaldi}
\affiliation{\pancaldi}

\author{Paolo Vavassori}
\affiliation{\nanogune}
\affiliation{\ikerbasque}
\email{p.vavassori@nanogune.eu}

\begin{abstract} 
	Multi-domain states of square artificial spin ice show a range of different morphologies ranging from simple stripe-like domains to more organically shaped coral domains.
	%
	%
	To model the relevant dynamics leading to the emergence of such diverse domain structures, simplified descriptions of the switching behavior of individual nanomagnets are necessary. 
	%
	In this work, we employ kinetic Monte Carlo simulations of the demagnetization of square artificial spin ice toward its ground state, and compare how the choice of transition barriers affect the emergence of mesoscale domains. 
	%
	We find that the commonly used mean-field barrier model (informed by equilibrium energetics only) results in propagation of ground-state string avalanches. In contrast, taking into account chiral barrier splitting enabled by state-dependent local torques supports the emergence of complex-shaped coral domains and their successful relaxation towards the ground state in later relaxation stages.
	%
	Our results highlight that intrinsic contributions to switching barriers, in addition to the effect of extrinsic defects often attributed to nanofabrication irregularities, can subtly shift favored transition pathways and result in different emergent mesoscale features. 
	%
    Future kinetic Monte Carlo models that describe the evolution of artificial spin systems should thus account for these effects.
\end{abstract}

\maketitle

%
Artificial spin ices (ASI) are metamaterials in which networks of interacting nanomagnets\cite{2006Wang, NISOLI2013,HEYDERMAN2013,2019Skjaervo} are used to study phenomena such as phase transitions and emergent correlations in geometrically frustrated spin systems\cite{MENGOTTI2008,MENGOTTI2011,BRANFORD2012,CANALS2016,SENDETSKYI2016} or using their self-organizing behavior for low power unconventional computing.\cite{IMRE2006,2018Arava,ARAVA2019,CARAVELLI2020,GARTSIDE2022}
A key question common to all these effects is how mutual interactions between the basic constituents (i.e., nanomagnets) impact the evolution of macroscopic variables, the emergence of mesoscale domains,\cite{MORGAN2011a,LEHMANN2020} and the successful relaxation to the ground state of the system.\cite{2006Wang, KAPAKLIS2012,FARHAN2013,Kapaklis2014}
Over the years, different mechanisms have been implemented to excite artificial spin systems, ranging from oscillating or rotating magnetic fields,\cite{WANG2007,BUDRIKIS2014,GILBERT2015,2021Bingham,BINGHAM2022} thermalization during sample growth,\cite{MORGAN2011a,NISOLI2012,2013Levis,2013Morgan} thermal annealing after growth,\cite{PORRO2013,ZHANG2013,ZHANG2019} as well as thermalization at constant temperature using superparamagnetic nanomagnets.\cite{KAPAKLIS2012,ANDERSSON2016,MORLEY2017,2019Sendetskyi,2019Chen_a,2020Arava,GORYCA2021,2022Goryca}

A well-studied specific lattice geometry is the square artificial spin ice (s-ASI), where four nanomagnets are arranged on cross-like vertices, as shown in Fig.~\ref{fig:foundations}(a).
The s-ASI has a well-defined antiferromagnetic ground state made up of a tiling of so-called $T_1$ vertices [Fig.~\ref{fig:foundations}(b)],\cite{2006Wang,2006Moller} which makes it attractive for comparative studies tracking its relaxation via successive moment reversals lowering the net interaction energy of the system.\cite{FARHAN2013, ARAVA2019}  

A particularly intriguing observation in relaxation studies relates to the emergent mesoscale structures imaged by magnetic microscopy, which can come in two distinct flavors: On the one hand, relaxation driven by avalanches of $T_1$ strings results in \textit{diagonal stripe-like domains} that clearly mirror the underlying energy {hierarchy} of the system.\cite{KAPAKLIS2012,2013Farhan,ANDERSSON2016,2020Arava,2021Bingham}
On the other hand, organically shaped \textit{coral} domains -- delimited by domain boundaries oriented along \textit{any} lattice directions (and thus not necessarily coinciding with the diagonal string propagation),\cite{MORGAN2011a, ZHANG2013, 2013Levis, PORRO2013, 2013Morgan, ZHANG2019} -- are usually attributed to extrinsic site-specific disorder originating from nanofabrication defects.\cite{HUGLI2012,BUDRIKIS2012,2013Farhan,GILBERT2015,JENSEN2002}

Supplementing experimental observations, Monte Carlo simulations based on simplified assumptions are often used to explain key features in the evolution of artificial spin systems.\cite{2012Silva,2013Farhan,ARAVA2019,JENSEN2022,2024Maes} 
Recently, we showed that chiral barrier splitting enabled by state-specific local torques has been overlooked so far in the description of switching behavior in s-ASI.\cite{KORALTAN2020, 2021Leo} Taking local fields into account, the switching barriers are significantly reduced in strongly interacting s-ASI, leading to faster relaxation as well as modifying favored relaxation pathways.

In this work, we expand on these findings on the demagnetization of s-ASI using kinetic Monte Carlo (kMC) simulations, comparing a model based on purely equilibrium energy considerations (mean-field model) with a model that takes into account the local contributions to the kinetic switching barriers. We find that -- even in the absence of disorder -- the modifications due to chiral barrier splitting can lead to the formation of coral domains, and identify the relevant relaxation pathways that favor the formation of complex-shaped domains as well as their final consolidation towards a global ground state. 
%
%
Our results demonstrate how the influence of intrinsic interactions, in addition to extrinsic disorder, can lead to the emergence of complex mesoscale domain patterns, an effect which should be considered for future modeling of relaxation of artificial spin systems.

\section{Background}

\begin{figure}[t!]
 	\includegraphics{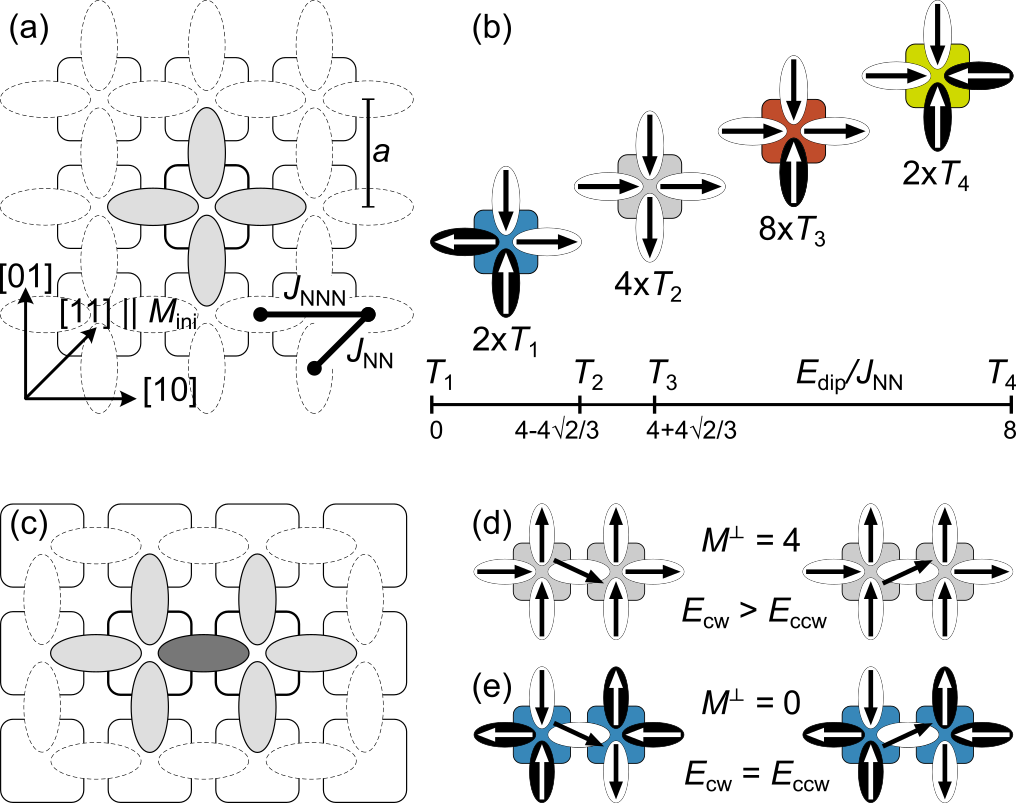}
	\caption{%
		\textbf{Square artificial spin ice.}
		(a)~Geometry of tiled four-moment vertices with lattice periodicity $a$ and nearest- and next-nearest neighbor interactions \JNN\ and \JNNN.
		(b)~Energy hierarchy and multiplicity of single-vertex configurations $T_i$ ($i=1\ldots4$).
		(c)~The double-vertex environment (light gray) acting on the central island (dark gray) determines the energy barrier for moment switching.
		(d,e)~For reversal via coherent rotation a perpendicular magnetization $M^\perp\ne 0$ exerted by the environment can lift the degeneracy for clockwise and counter-clockwise reversal energy barriers. 
        }
	\label{fig:foundations}
\end{figure}


Here, we consider the s-ASI lattice with periodic boundary conditions, which is formed by $n\times n$ four-moment vertices (i.e., in total $2n^2$ moments), as shown in Fig.~\ref{fig:foundations}(a). 
Due to rapid interaction drop-off, it is sufficient to consider interactions between nearest and next-nearest neighbors.\cite{2012Silva,2013Morgan,2013Farhan,2021Leo}
Within the point-dipole approximation, interaction strengths \JNN\ and \JNNN\ depend on the lattice periodicity $a$ and magnetic moment $m$ only, with $\mu_0$ being the magnetic permeability:
\begin{eqnarray}
	\JNN &=  \frac{3}{\sqrt{2}} &\frac{\mu_0}{2\pi} \frac{m^2}{a^3} \quad \text{and} \label{eq:JNN} \\
	\JNNN &= &\frac{\mu_0}{2\pi} \frac{m^2}{a^3} = \frac{\sqrt{2}}{3} \JNN \approx 0.47\JNN \,. \label{eq:JNNN}
\end{eqnarray}

The 16 possible states of a four-moment vertex define an energy hierarchy with four distinct cases, Fig.~\ref{fig:foundations}(b). The two lowest-energy states are ground-state $T_1$ configurations (blue), and $T_2$ vertices (gray) that can be initialized by a diagonal saturating magnetic field. The $T_1$ and $T_2$ configurations obey the so-called two-in-two-out "ice rule",\cite{2006Wang} whereas $T_3$ (red) and $T_4$ (yellow) are high-energy configurations associated with magnetic charges. $T_4$ configurations are especially unfavorable and rarely occur in experiments or simulations.\cite{MORGAN2011a,2012Phatak}


In this work, we consider the thermal demagnetization from a field-set initial state of $T_2$ vertices to a low energy configuration dominated by $T_1$ vertices.\cite{2013Farhan, ANDERSSON2016,2021Bingham} Full demagnetization requires every other moment to flip, and thus at least $n^2$ spin flips to reach the ground state. However, any transition from a $T_2$ to a $T_1$ state requires the excitation of higher-energy $T_3$ vertices, as mutual interactions restrict the relaxation pathways.\cite{FARHAN2013,ARAVA2019}

To model the time-dependent demagnetization, the transition rates $\nu(\Delta E, T)$ for thermally activated moment reversals can be expressed by the Arrhenius law: 
\begin{equation}
	\nu(\Delta E, T) = \nu_0 \exp\left(-\frac{\Delta E}{\kB T} \right) \, .
	\label{eq:rate:Arrhenius}
\end{equation}
Here, $\Delta E$ is the energy barrier to be overcome, $T$ is the absolute temperature and $\kB$ the Boltzmann constant.\cite{BROWN1963, 1969Langer} The prefactor $\nu_0$ is a rate typically in the order of \num{e8}-\SI{e12}{s^{-1}} and depends on the size and temperature of the nanomagnet.\cite{BROWN1963,KRAUSE2009,2018Stier} Here, we assume that $\nu_0$ is state-independent, or at least varies much less than the exponential dependence with the barrier energy.\cite{DESPLAT2020,2021Kanai}

As shown in Fig.~1(c), a moment surrounded by six closest neighbors forms a double-vertex configuration that determines the energy contribution to the switching barrier $\Delta E$. For s-ASI in the limit of quasi-coherent moment reversal, $\Delta E$ can be decomposed into three contributions.\cite{KORALTAN2020,2021Leo}

First, the single-particle barrier $\Delta E_\text{sb}$ is associated with the shape anisotropy of a non-interacting magnetic nanoisland.\cite{OSBORN1945,CHEN1991,COWBURN2000,IMRE2006} In this work, we use $\tau_{sb}$ as a natural time scale of the evolution: 
\begin{equation}
	\tau_\text{sb}^{-1} = \nu(\Delta E_\text{sb}, T) \, .
	\label{eq:rate:single}
\end{equation}

Second, the single-particle barrier is modified by the pairwise interactions and favors transitions to configurations of lower dipolar energy. Using the energy difference $\Delta E_{i\rightarrow f}^\text{dip} = E_f^\text{dip} - E_i^\text{dip}$ between final and initial static moment configurations, the "mean-field" (MF)\cite{FARHAN2013,2018Arava,ARAVA2019, KORALTAN2020, 2021Leo} switching rate is given by
\begin{equation}
	\nu_i^\text{MF} = 2 \nu_0 \exp\left(-\frac{\Delta E_\text{sb} + \frac{1}{2} \Delta E_{i\rightarrow f}^\text{dip} }{\kB T} \right) \, .
	\label{eq:rate:meanfield}
\end{equation}

Third, as shown in Figs.~\ref{fig:foundations}(d,e), the four nearest-neighbor moments can exert a torque via a local perpendicular net magnetization $M^\perp_i$ acting on the switching moment. If $M^\perp_i\ne0$, the energy barriers for clockwise (\cw) and counter-clockwise (\ccw) switching differ:\cite{KORALTAN2020, 2021Leo}
\begin{equation}
	\Delta E_{i\rightarrow f,\text{\cw/\ccw}} = \Delta E_\text{sb} + \frac{1}{2}\Delta E_{i\rightarrow f}^\text{dip} \pm \frac{M^\perp_i}{3}\JNN \, .
	\label{eq:barrier:chiral}
\end{equation}
This chiral barrier splitting is present for a large fraction of magnetic configurations,\cite{2021Leo} including the fully-magnetized background [Fig.~\ref{fig:foundations}(d)] where, with $M_\perp=+4$, a counter-clockwise reversal is favorable over a clockwise moment rotation. In contrast, barrier splitting is absent for example for the ground-state configuration with $M_\perp=0$ [Fig.~\ref{fig:foundations}(e)].

Due to the exponential behavior of the Arrhenius law in Eq.~(\ref{eq:rate:Arrhenius}), even a small reduction of the energy barrier significantly increases the switching rate through the favored (clockwise or counterclockwise) reversal channel. In the following we denote the increased net relaxation rate as "chiral-split barrier model", or CB:
\begin{equation}
	\nu_i^\text{CB} =\nu_i^\text{\cw}+\nu_i^\text{\ccw}
	= \nu_i^\text{MF} \cosh\left( M^\perp_i\frac{\JNN}{3 \kB T}  \right)
	\label{eq:rate:chiral}
\end{equation}

\section{Methods}

We performed kinetic Monte Carlo (kMC) simulations\cite{1975Bortz,1991Fichthorn,1999Newman} of the demagnetization of s-ASI, using a custom-written \texttt{python} kMC code \cite{Matteo-github, 2018Pancaldi}, comparing the relaxation behavior of mean-field and chiral-split barrier models.\cite{2021Leo, KORALTAN2020}.  
%
%
Periodic boundary conditions ensure that all moments have the same double-vertex environment, emulating a larger network without edge effects. In particular, we assume a perfect system with equal moments, that is, we do not take into account the effect of defects or site-specific variations in switching barriers. The purely configuration-dependent probabilities for successive spin flips are calculated from the current state and using a lookup table for the double-vertex transition rates calculated from Eq.~(\ref{eq:rate:meanfield}) and Eq.~(\ref{eq:rate:chiral}), respectively. For each kMC step, a single moment is flipped after being randomly selected from the weighted transition probabilities, and, together with a random time update (again, based on the most likely transition rates), the system is updated.\cite{1975Bortz,1991Fichthorn,1999Newman, 2018Pancaldi}

The single-particle switching barrier $\Delta E_\text{sb}=\SI{1.327}{eV}$ and moment $m=\num{3.285e06}\mu_\mathrm{B}$ correspond to values for a stadium-shaped permalloy island with length, width and height of $\SI{150}{nm}\times\SI{100}{nm}\times\SI{3}{nm}$ and a saturation magnetization of \SI{790}{kA/m}. As shown in previous works\cite{KORALTAN2020, 2021Leo} such a magnetic nanoisland reverses via quasi-uniform modes and thus is an appropriate choice for our study.\cite{MORLEY2017,BINGHAM2022} 

The lattice periodicity $a$, and thus the strength of \JNN\ in Eq.~(\ref{eq:JNN}), has been varied between values of \SI{150}{nm} and \SI{500}{nm}. The resulting Néel temperatures $\TN(a)=1.7\JNN(a)$\cite{2006Moller,2012Silva} is always above the fixed simulation temperature of $T=\SI{300}{K}$ with ratios $T_\mathrm{N}/T$ ranging from 1.3 (i.e., close to the transition to the paramagnetic phase) to $\approx 30$ (i.e., $T\rightarrow 0$) [see Supplemental Materials Fig.~\ref{sfig:TN-vs-a}].

%

We chose to simulate a s-ASI with a system size of $50\times 50$ vertices ($2n^2=5000$ moments), compromising between statistics and computational resources [see Supplementary Materials Fig.~\ref{sfig:system_size_comparison}]. 
The initial configuration is fully magnetized with $M_\text{ini}\parallel[11]$ parallel to the diagonal $[11]$ direction [see Fig.~\ref{fig:foundations}(a)], and the system is given $n^3=\num{125e3}$ steps to relax. For each value of $a$ and the barrier model, at least \num{20} kMC runs have been executed to ensure meaningful statistics.

\begin{figure}[t!]
	\includegraphics{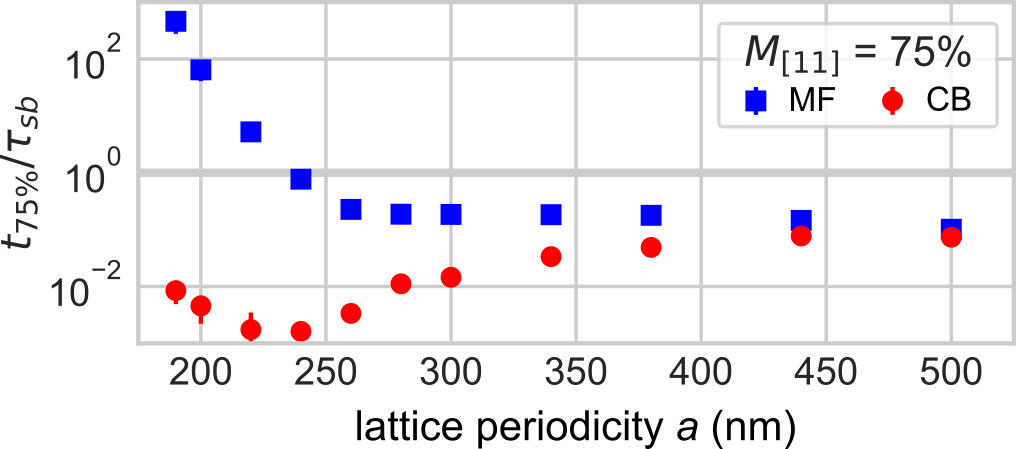}
	\caption{%
		\textbf{Average time at which $M_{[11]}$=75\%} for the mean-field barrier model (blue squares) and the chiral-split barrier model (red circles). 
		Both models approach each other at large lattice periodicities $a$.
		For strong interactions (small lattice periodicity $a$), the mean-field model shows slower evolution compared to the chiral-barrier model, as the latter generally features faster kinetics due to the reduction of switching barriers. 
		For periodicities $a\le\SI{240}{nm}$ both models show an apparent slowdown. 
		For the mean-field model, this slowdown is even beyond the single-particle relaxation time scale $\tau_\mathrm{sb}$ due to string propagation being the only significant relaxation pathway. 
		For the split-barrier model, the slowdown originates from the effect of kinetic blocking via subsequent creation and annihilation of $T_3$ vertices.
	}
	\label{fig:quantile_times}
\end{figure}

\begin{figure*}[t!]
	\includegraphics{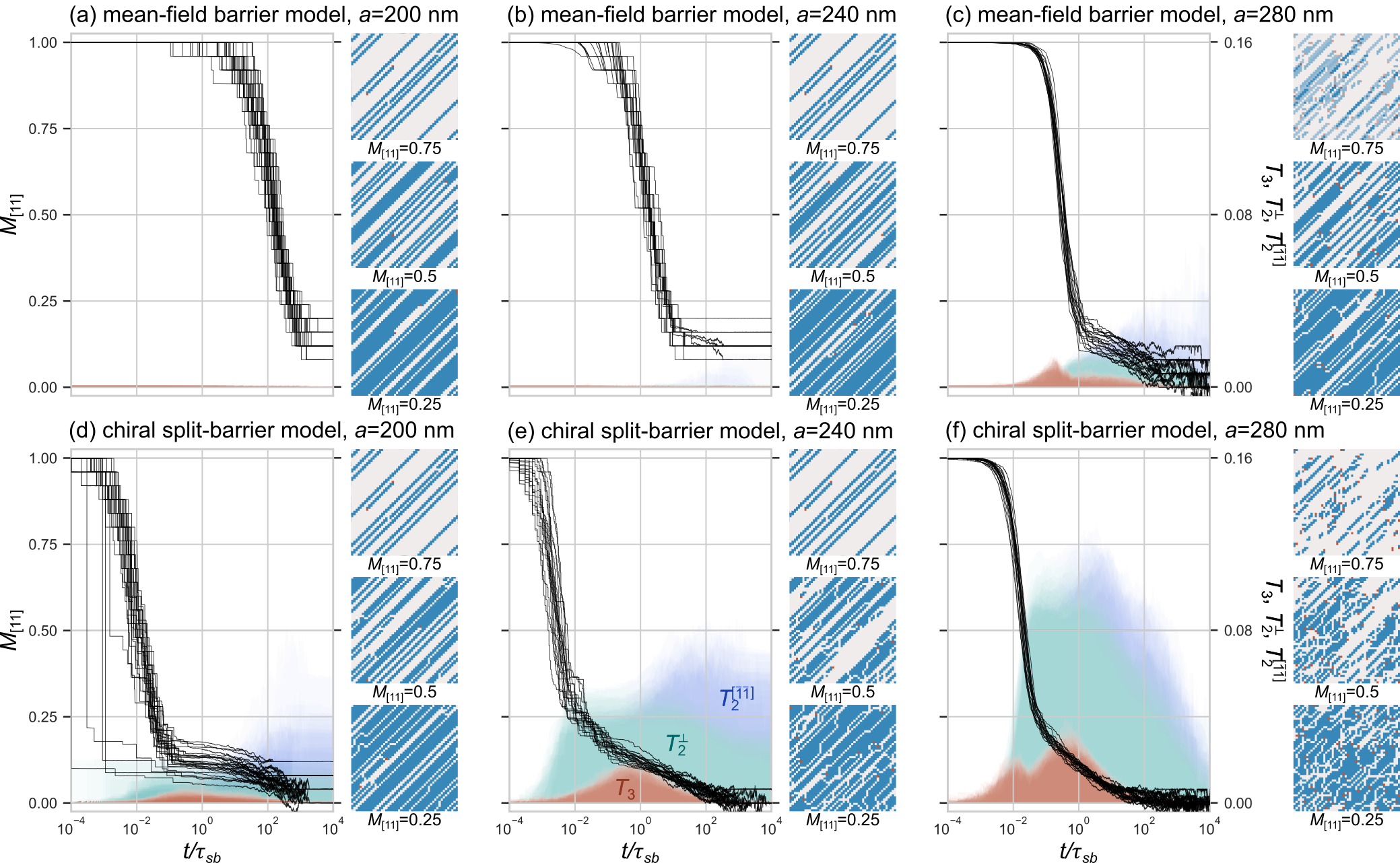}
	\caption{%
		\textbf{Demagnetization behavior for different lattice periodicities $a$} for (a-c)~the mean-field barrier model and (d-f)~the chiral-split barrier model.
		Black lines show the time evolution of the net magnetization $M_{[11]}(t)$ of different kMC runs (left-handed scale from 0~to~1).
		Shaded areas (right-handed scale from 0~to~\num{0.16}) denote the population of $T_3$ vertices (red, in forefront) and $T_2$ vertices that do not align with the initial $[11]$ direction, separated into rotated $T_2^\perp$ (cyan, from zero) and reversed $T_2^{[\bar{1}\bar{1}]}$ (blue, stacked on $T_2^\perp$ population). 
		Insets on the right show representative spatial configurations of vertices at times where the net magnetization is at 75\%, 50\% and 25\% of its initial value. Vertex color coding is equivalent to Fig.~\ref{fig:foundations}(a). 
	}
	\label{fig:demagnetisation}
\end{figure*}

\section{Results}


To compare the relaxation time scales, Fig.~\ref{fig:quantile_times} shows the mean time $\left<t(M_{[11]}=75\%)\right>_\text{runs}$, normalised to $\tau_\mathrm{sb}$. For the mean-field model [blue squares] we find that with strong interactions \JNN\ (respectively, lattice periodicities $a\le\SI{240}{nm}$) the speed of initial demagnetization are slowed down even beyond the single-moment switching time.
This is in contrast to the chiral-split barrier model [red circles] for which the demagnetization evolves significantly faster but slows down with increasing lattice periodicity $a$ and approaches the mean-field model. This convergence arises from the fact that the interaction strength $\JNN\propto a^{-3}$ rapidly decreases and the transition barriers are dominated by $\Delta E_\text{sb}$ only (alternatively, $\TN(a)\rightarrow T$, see Supplemental Materials Figs.~\ref{sfig:TN-vs-a} and \ref{sfig:demagnetisation-large_a}).
%
For small $a\le\SI{240}{nm}$ the relaxation in both models slows down at different rates (due to reasons discussed in the next section), with the relaxation time scales for mean-field barrier model exceeding that of the single-particle relaxation time scale given by $\tau_\mathrm{sb}$. 


Figure~\ref{fig:demagnetisation} shows kMC results of the demagnetization process in s-ASI with lattice periodicities $a=\SI{200}{nm}$, \SI{240}{nm} and \SI{280}{nm} (from left to right), comparing the behavior of the mean-field barrier model [(a-c), top row] to that of the chiral-split barrier model [(d-f), bottom row].
%
The black lines track the net magnetization $M_{[11]}(t/\tau_\text{sb})$ for each kMC run (left scale) with the time $t$ normalized to the switching time scale $\tau_{sb}$ of a non-interacting moment defined in Eq.~(\ref{eq:rate:single}).

The magnetization $M_{[11]}(t)$ decreases on an exponential time scale to a value of around 25\% of the initial moment, after which the relaxation slows down considerably. 
%
Especially in the case of the chiral-split barrier model, the evolution can continue to a value of $M_{[11]}(t)\approx 0$, which, however, does \textit{not} imply that a fully relaxed $T_1$ ground state has been reached. For the mean-field model, full demagnetization is less often observed due to a limited simulation time window, as well as kinetic blocking of effective relaxation in later stages (discussed in Sec.~\ref{sec:domain_morphologies_emergence}).

%
It is instructive to also consider the vertex populations of $T_3$ (red). 
For lattice periodicities $a\le\SI{250}{nm}$ the initial demagnetization from $M_{[11]}$=100\% to 50\% is accompanied by a small stable population of $T_3$ vertices ($\le$0.5\%, i.e., corresponding to at most 14 vertices in our system), indicating correlated demagnetization.
At higher values of $a$, respectively weaker interactions, the $T_3$ population peaks at a time where $M_{[11]}\approx$75\%, indicating more randomized demagnetization, as shown in Figs.~\ref{fig:demagnetisation}(c) and (f) as well as Fig.~\ref{sfig:demagnetisation-large_a}. 
A later peak in $T_3$ vertices around $M_{[11]}\approx$10\% is associated with dynamics confined to ground-state domain boundaries, with further details discussed in Sec.~\ref{sec:transitions_at_domain_boundaries}.

Further insights into the relaxation behavior can be obtained from representative real-space snapshots of the magnetic configuration at $M_{[11]}$=75\%, 50\% and 25\%, shown in the three panels on the right side of each graph, with vertex color coding according to Fig.~\ref{fig:foundations}(b).
For the mean-field model, Fig.~\ref{fig:demagnetisation}(a-c), the snapshots are dominated by diagonal lines of $T_1$ vertices (blue), related to the avalanche-like propagation of $T_1$ "strings" through the fully magnetized background (gray).\cite{MORGAN2011,2013Farhan,2020Arava,2021Bingham}  
For the chiral-split barrier model, Fig.~\ref{fig:demagnetisation}(d-f), the picture is more diverse, with more $T_3$ vertices present and $T_1$ strings broken up by sideways interruptions, and at $M_{[11]}$=25\% (lower panels) indications for the emergence of coral-shaped ground state domains.\cite{MORGAN2011a,ZHANG2013,PORRO2013}

\begin{figure*}[t!]
	\centering
    \includegraphics{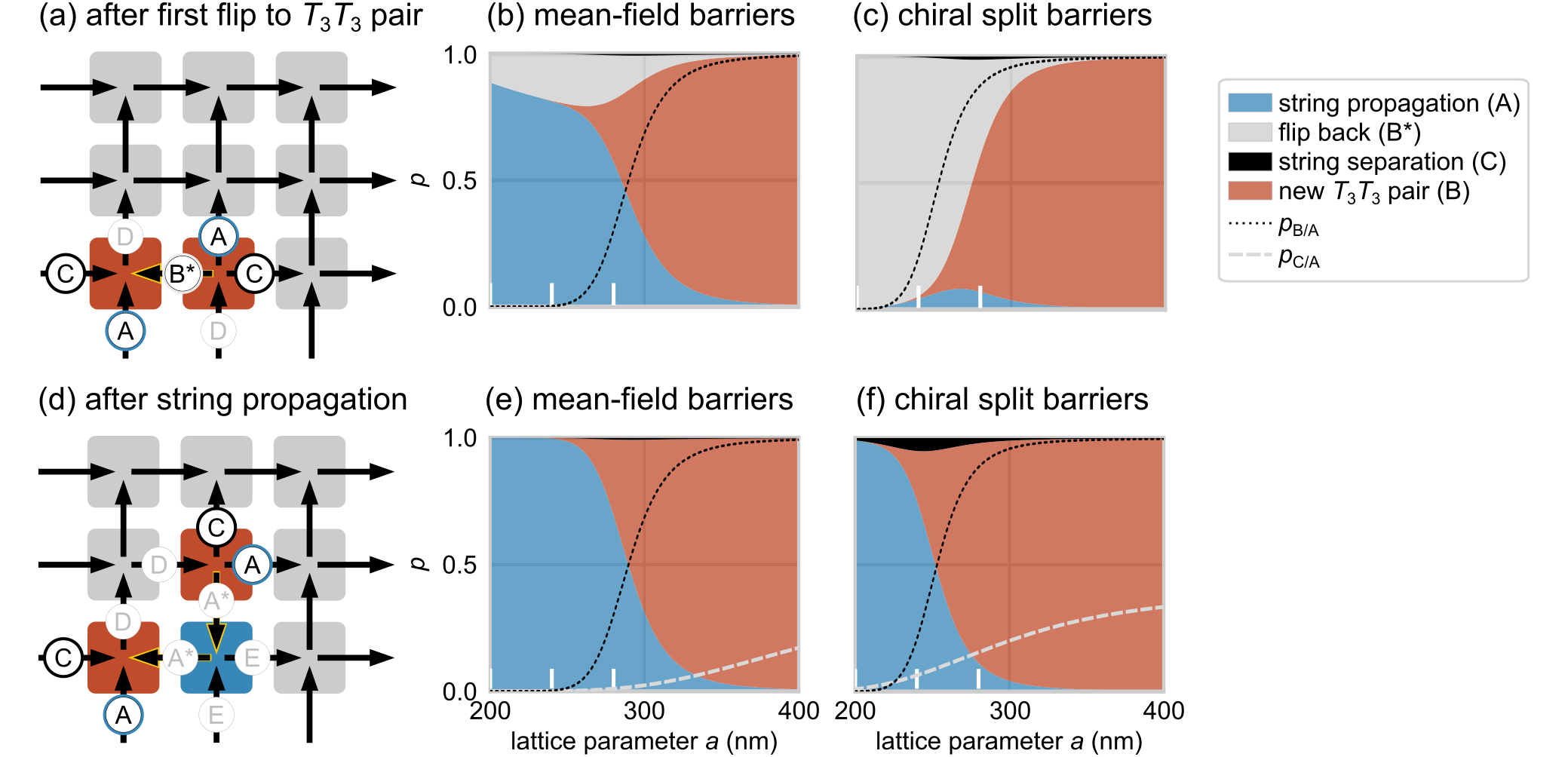}
	\caption{%
		\textbf{Initial demagnetization steps.}
		(a)~Spin configuration after a first spin flip creating a $T_3$ pair. Distinct transitions are labeled with letters, and unmarked moments are of transition type B ($T_2T_2\rightarrow T_3T_3$, see Tab.~\ref{tab:transitions}). Transitions which are highly unlikely to occur are marked in light gray (e.g., transition D).
		(b,c)~Transition probabilities $p_i$ in dependence of the lattice periodicity $a$ for (b)~mean-field barrier energies and (c)~when taking into account additional chiral splitting of the barriers. The probability $p_\text{B/A}$ (black dotted line) demarks the transition between demagnetization dominated via string propagation (transition A, blue) and creation of new $T_3$ pairs (B, red). Flip back transitions that restore the initial fully-magnetized $T_2$ state can be highly probable (B*, gray), and, in the case of the split-barrier model, are favored for strong interactions [i.e., smaller $a$ in (c)]. 
		White markers indicate the lattice periodicities $a$ shown in Fig.~\ref{fig:demagnetisation}.
		(d)~Spin configuration and labelled transition after the formation of a first $T_1$ vertex (proto-string), which is a stable emerging magnetic texture.
		(e,f)~Comparison of model-dependent transition probabilities. The probability $p_{C/A}$ (dashed gray line) indicates the ratio of string separation (transition C) over string propagation events (transition A). String separation events result in $T_2^\perp$ vertices important for the formation of coral-shaped domain states, and are more likely to occur for the split-barrier model.
		}
	\label{fig:relative_rates_initial_steps}
\end{figure*}

\begin{table}[tb]
	\caption{%
		\textbf{Transition barriers for Figs.~\ref{fig:relative_rates_initial_steps} and \ref{fig:domains}}, labeled A to L.
		For each panel, possible transitions are sorted by increasing mean-field barrier $\Delta E^\text{dip}_{i\rightarrow f}$, respectively, decreasing transition rate. The barrier energies are both given in multiples of \JNN\ and \JNNN\ (left), as well as approximate multiples of \JNN\ (right). 
		The value of $M_i^\perp$ indicates whether the switching macrospin is subject to a torque by its magnetic neighborhood, a condition that enables chiral barrier splitting. 
		The last column denotes the number of possible sites $N_i$ for each spin flip shown in Figs.~\ref{fig:relative_rates_initial_steps}~(a) and (d), which are necessary to calculate initial transition probabilities.
		}\vspace{1ex}
	\label{tab:transitions}
	\centering
	\begin{tabular}{l|c|lr|r|c}
		& transition & \multicolumn{2}{c|}{$\Delta E_{i\rightarrow f}^\text{dip}$} & \multicolumn{1}{c|}{$M_i^\perp$} & $N_i$\\ 
		\hline 
		
		\multicolumn{6}{l}{\textbf{Fig.~\ref{fig:relative_rates_initial_steps}(a): Transitions after initial $T_3T_3$ pair}}\\
		A & $T_3T_2\rightarrow T_1T_3$  &  $-4\JNN+4\JNNN$ & $ -2.11$  & $\pm2$ & $2$ \\
		B* & $T_2T_2\leftarrow T_3T_3$ &  $-4\JNNN$ & $-1.89$ & $\pm2$ & $1$\\
		C & $T_3T_2\rightarrow T_2^\perp T_3$  & $0$ & $0$ & $\pm4$ & $2$\\
		B & $T_2T_2\rightarrow T_3T_3$ & $+4\JNNN$ & $+1.89$ & $\pm4$ & $2n^2-7$\\
		D & $T_3T_2\rightarrow T_4T_3$ & $+4\JNN+4\JNNN$ & $+5.89$  & $\pm2$ & $2$\\
		\hline
		
		\multicolumn{6}{l}{\textbf{Fig.~\ref{fig:relative_rates_initial_steps}(b): Transitions after string formation}}\\
		A & $T_3T_2\rightarrow T_1T_3$  & $-4\JNN+4\JNNN$ & $ -2.11$  & $\pm2$ & $2$\\
		C & $T_3T_2\rightarrow T_2^\perp T_3$ & $0$ & $0$ & $\pm4$ & $2$\\
		B & $T_2T_2\rightarrow T_3T_3$ & $+4\JNNN$ & $+1.89$ & $\pm2$ & $2n^2-10$ \\
		A* & $T_3T_2\leftarrow T_1T_3$ & $+4\JNN-4\JNNN\quad$ & $ +2.11$  & $\pm2$ & $2$\\
		E & $T_1T_2\rightarrow T_3T_3$& $+4\JNN$  & $+4$ &  $\pm2$& $2$ \\
		D & $T_2T_3\rightarrow T_4T_3$  & $+4\JNN+4\JNNN$ & $+5.89$  & $\pm2$ & $2$\\
		\hline

		\multicolumn{6}{l}{\textbf{Fig.~\ref{fig:domains}: Transitions at domain boundaries}}\\
		F* & $T_1T_2\leftarrow T_3T_3$  & $-4\JNN$ & $-4$ & $\pm2$ &\\
		A & $T_3T_2\rightarrow T_1T_3$  & $-4\JNN+4\JNNN$ & $-2.11$ & $\pm2$&\\
		G* & $T_3T_3\rightarrow T_2T_2$  & $-4\JNNN$ & $-1.89$ & $0$&\\
		H & $T_3T_1\leftrightarrow T_1T_3$  & $0$ & $0$ & $0$&\\
		J & $T_2T_3\leftrightarrow T_3T_2$  & $0$ & $0$ & $0$&\\
		G & $T_3T_3\leftarrow T_2T_2$  & $+4\JNNN$ & $+1.89$ & $0$&\\
		A* & $T_3T_2\leftarrow T_1T_3$  & $+4\JNN-4\JNNN$ & $+2.11$ & $\pm2$&\\
		F & $T_1T_2\rightarrow T_3T_3$  & $+4\JNN$ & $+4$ & $\pm2$ &\\
		K & $T_1T_1\rightarrow T_3T_3$  & $+8\JNN-4\JNNN$ & $+6.11$ & $0$&\\
		L & $T_1T_3\rightarrow T_3T_4$  & $+8\JNN$ & $+8$ & $0$&\\
	\end{tabular}
\end{table}

\subsection{Initial Demagnetization: Pathways}
\label{sec:init_demag_patways}

To understand the preferred switching pathways and relaxation regimes in Fig.~\ref{fig:relative_rates_initial_steps} we compare the relative transition rates for the initial demagnetization steps of the s-ASI.
%
Fig.~\ref{fig:relative_rates_initial_steps}(a) shows the spin configuration after a first spin flip on a random site within the magnetized $T_2$ background, which creates a pair of higher-energy $T_3$ vertices. Possible next transitions are labeled with letters A to D.
%
Transition A creates a ground-state $T_1$ vertex along the $[11]$ diagonal. In contrast, transition D would create a $T_4$ vertex and thus is unlikely to occur (dimmed gray label). The imbalance between transitions A and D drives the distinct demagnetization parallel with $M_\text{ini}$ and the emergence of $T_1$ strings.\cite{2013Farhan,2020Arava,2021Bingham} Further possible transitions are the creation of an independent $T_3$ pair (transition B, not labeled explicitly), a flip-back to the original saturated $T_2$ state (transition B*), or the sideways separation of the $T_3$ pair (transition C).

The probabilities $p_i$ for possible transitions $i$=(A, B, B*, C, D) can be calculated according to 
\begin{equation}
	p_i^\text{model} = \frac{N_i \nu_i^\text{model} }{\sum_j N_j \nu_j^\text{model}} \, .
	\label{eq:relative_probability}
\end{equation}
The transition rates $\nu_i^\text{model}(a,T)$ for the mean-field [chiral-split] barrier model can be calculated from Eq.~(\ref{eq:rate:meanfield}) [Eq.~(\ref{eq:rate:chiral})], using the value of $\JNN\propto a^{-3}$ according to Eq.~(\ref{eq:JNN}). The corresponding energy differences $\Delta E_{i\rightarrow f}^\text{dip}$, perpendicular magnetization $M_i^\perp$ and number of possible sites $N_i$ after the first spin flips are listed in Tab.~\ref{tab:transitions}.

Figs.~\ref{fig:relative_rates_initial_steps}(b) and (c) show stacked plots of the relative transition probabilities in dependence of the lattice constant $a$ for the mean-field and chiral-split barrier model, respectively.
One major difference between both models is the backflip probability (transitions B*) after creation of a $T_3$ pair (gray area). In the case of the chiral-split barrier model and for strong interactions $\JNN\propto a^{-3}$  transition B* is the preferred relaxation pathway, due to the reduction of the switching barrier enabled by high local torques. In a defect-free system (as is the case in our simulations), the constant creation and annihilation of $T_3$ pairs (transitions B*) creates a situation of "kinetic blocking", which delays the onset of the demagnetization via $T_1$ string propagation [see Fig.~\ref{fig:quantile_times}].
 
Apart from the peculiarity of back-flips boosted by chiral barrier splitting, the two main pathways toward efficient demagnetization are either the string propagation (transition A, blue) or the creation of new $T_3$ pairs (transition B, red). The former transition is energetically preferred as it creates a ground state $T_1$ vertex, whereas the probability for the latter is enhanced by the $2n^2-7$ sites in which the transition can take place (see Tab.~\ref{tab:transitions}).
The relative probability $p_\text{B/A}$ of these two transitions can be calculated as
\begin{equation}
    p_\text{B/A}^\text{model} = \frac{N_B \nu_B^\text{model} }{N_A \nu_A^\text{model} + N_B \nu_B^\text{model}} \, ,
    \label{eq:pB_A}
\end{equation}
and is shown as dotted black line in Figs.~\ref{fig:relative_rates_initial_steps}(b) and (c). This line demarks the range of lattice periodicity $a$ where one relaxation pathway is favoured over the other. 
In contrast, for the chiral-split barrier model, demagnetization via $T_3$ pair formation becomes favoured over string propagation at smaller values of $a$, compared to the mean-field barrier model. 

Once a $T_1$-string is nucleated, Fig.~\ref{fig:relative_rates_initial_steps}(d), the relative importance of the relaxation pathways shifts dramatically, see Figs.~\ref{fig:relative_rates_initial_steps}(e,f). In particular, the $T_1$ vertices enjoy remarkable stability, with back-flip A* or transitions D and E extremely unlikely. 
For the mean-field barrier model and low values of $a\le\SI{240}{nm}$, string propagation remains the \textit{only} available relaxation pathway. Due to the way the kMC simulations are implemented,\cite{Matteo-github, 2018Pancaldi} the update time on each step is dominated by the much larger energy barrier needed to form a new $T_3$ pair on the approximately $\approx 2n^2-m$ sites (with $m$ being the string length), leading to the exceptional slowdown in Fig.~\ref{fig:quantile_times}.

In general, the main evolution of the $T_1$ string occurs via diagonal $T_3$ propagation creating more $T_1$ vertices (transition A, blue) or the sideways separation of $T_3$ vertices (C, black), creating a $T_2^\perp$ vertex whose moments points perpendicular to $M_\mathrm{ini}$. In Figs.~\ref{fig:relative_rates_initial_steps}(e,f) the relative probability 
\begin{equation}
    p_\text{C/A}^\text{model} = \frac{N_C \nu_C^\text{model} }{N_A \nu_A^\text{model} + N_C \nu_C^\text{model}} \, ,
    \label{eq:pB_A}
\end{equation}
of the C and A transitions are shown as dashed gray lines.
The C transition, with $T_3T_2\rightarrow T_2^\perp T_3$, does not change the dipolar energy of the spin configuration, but -- due to $|M_i^\perp|=4$ -- is subject to strong chiral barrier splitting, which significantly enhances its probability compared to the mean-field barrier model at the same value of $a$. 
Typically, several $T_1$ strings grow during the initial demagnetization phase, and as such string separation events can lead to a sizable prevalence of $T_2^\perp$ vertices, which is also indicated by the cyan-shaded areas in Fig.~\ref{fig:demagnetisation}(d-f) and the larger number of breaks in $T_1$ strings within the corresponding snapshots.

\begin{figure*}[t!]
	\centering
	\includegraphics[width=\textwidth]{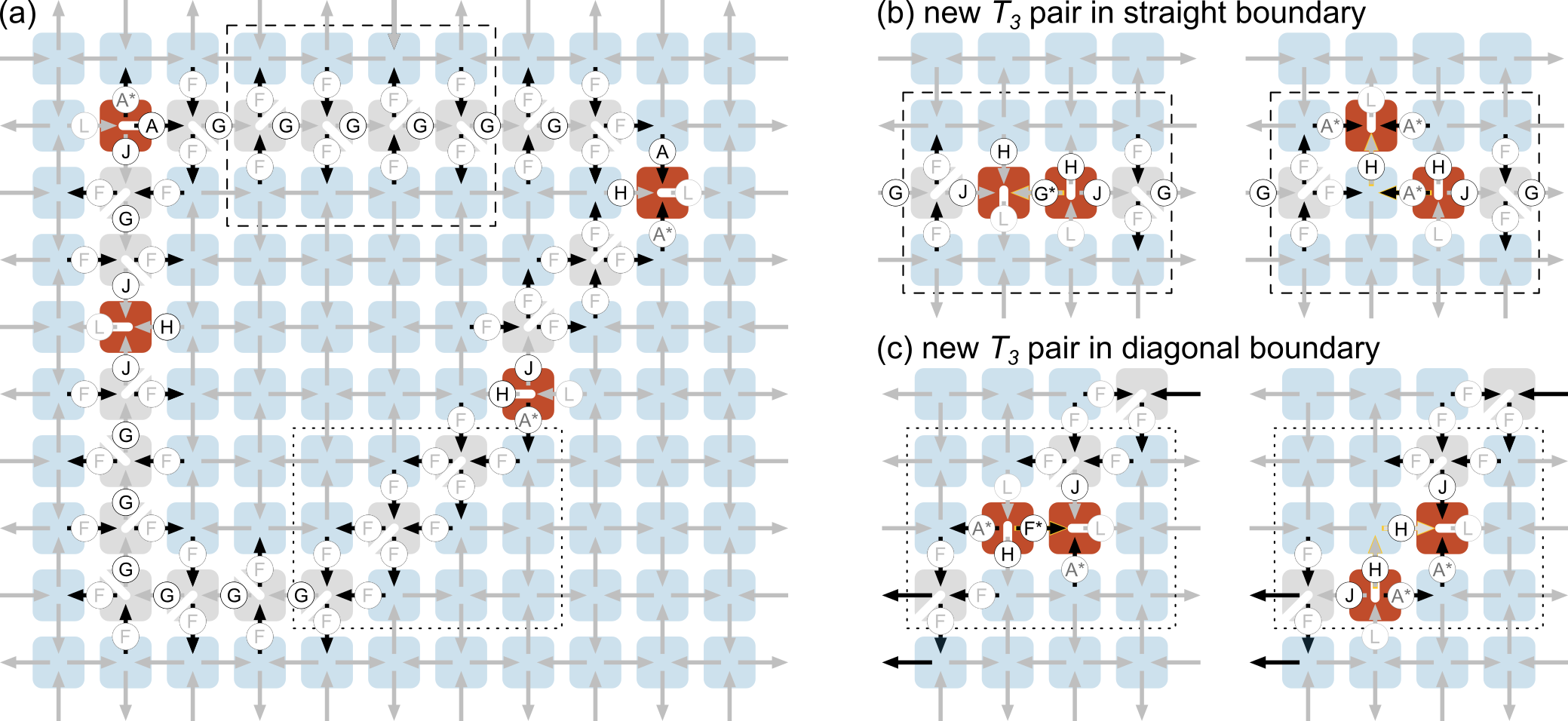}
	\caption{%
		\textbf{Transitions at domain boundaries}.
		Black arrows denote moments within local environments with $M^\perp\ne0$ that enable chiral barrier splitting, and thus can lead to enhanced transition rates. Gray arrows conversely indicate environments with $M^\perp=0$.
		Distinct transitions at the domain boundary are labeled, with gray labels indicating those with a high energy barriers and thus low probability. Transitions on any unlabeled spin are high-energy excitation $T_1T_1\rightarrow T_3T_3$ of the ground-state background which are unlikely to occur (transition type K in Tab.~\ref{tab:transitions}). 
		(a)~Prototypical domain boundary separating degenerate ground states. The boundary carries a net magnetization (indicated with white lines) that follows the outline of the domain. Most switching activity will involve $T_3$ defects located within the boundary.
		(b,c)~Following the creation of $T_3$ pairs in (b)~straight and (c)~diagonal domain boundaries, transition rates feature an asymmetry with respect to both ground-state domains on either side of the boundary, driving the deformation and shrinking of domains.
	} 
	\label{fig:domains}
\end{figure*}

\subsection{Generalised Results}

The kMC simulation results presented above have been obtained for specific values of the lattice periodicity $a$ and temperature $T$, while keeping the single-island magnetic moment $m$ and switching barrier $\Delta E_\mathrm{sb}$ fixed. 
Variation of any of these parameters strongly affect the switching rates calculated from Eqs.~(\ref{eq:rate:meanfield}) and (\ref{eq:rate:chiral}), and, in turn, lead to vastly modified relaxation pathways. A direct comparison of simulation outcomes that differ in more than one parameter therefore requires the consideration of reduced energy or temperature scales: 

Using the ratio $\TN/T\propto\JNN/T$ to denote the relative strength of interactions compared to the thermal energy, the transition probabilities are equivalent when varying the lattice periodicity $a$ or temperature $T$ [see Supplemental Material Fig.~\ref{sfig:initsteps_a_vs_T}], as $\JNN/T$ is the relevant factor modifying the energy barrier in the Arrhenius law. 

For a series of kMC simulations with varying values of $a$ and $T$ [Supplemental Material Fig.~\ref{sfig:phase_diagram}], we furthermore find that for, a given model, and normalised energies $\TN/T$ and $\num{20}\le\Delta E_\mathrm{sb}/(\kB T)\le\num{80}$ the simulation results are largely equivalent. 
Ratios $\Delta E_\mathrm{sb}/(\kB T)\ge\num{20}$ are consistent with values typical for meta-stable or slowly fluctuating nanomagnets imaged in x-ray photoelectron emission (PEEM) or magnetic force microscopy (MFM) experiments. \cite{MENGOTTI2008, MENGOTTI2011, MORGAN2011a, BRANFORD2012, 2013Farhan, CANALS2016, 2018Arava,ARAVA2019, LEHMANN2020}
%

In general, the mean-field and the split-barrier models exhibit distinct relaxation behaviour, as discussed above, except when approaching the paramagnetic regime with $T\rightarrow\TN$ where the relative influence of the chiral barrier splitting on the relaxation rates vanishes.\cite{2021Leo} 

\subsection{Emergence of Domain Morphologies}
\label{sec:domain_morphologies_emergence}

The favoured transitions A to D discussed above remain remarkably similar within the first three quantiles of relaxation, i.e., as $M_{[11]}$ decreases from 100\% to 25\% (which is typically achieved within a relatively small number of individual transitions).
As $T_3$ pairs are created and $T_1$ strings grow, two distinct domain morphologies might emerge, which are \textit{diagonal stripe domains} and \textit{coral domains} [see 25\% snapshots in Fig.~\ref{fig:demagnetisation}]. 

%
\textit{Stripe domains} are predominant for the mean-field model with strong interactions [e.g., Fig.~\ref{fig:demagnetisation}(a,b)]. As $T_1$ strings can form independently, the relative position of the lines can match to form extended areas of $T_1$ tiling. Alternatively, if $T_1$ strings are separated by a single diagonal of $T_2$ vertices, this boundary cannot be removed from the system without the generation of high-energy $T_4$ vertices. This very stable configuration (in addition to the limited simulation steps) makes domain consolidation -- that is, domain merging and annihilation -- toward a global ground state difficult, and leads to a stagnation of relaxation at final magnetization values around $M_{[11]}\approx$ 10-20\% [Fig.~\ref{fig:demagnetisation}(a,b)]. 

\textit{Coral domain} mainly appear for an interaction regime where demagnetization is neither dominated by string propagation nor by random creation of $T_3$ pairs, i.e, at intermediate values for $p_{B/A}$ [dotted black line in Figs.~\ref{fig:relative_rates_initial_steps}(e,f)], and where a sizable fraction of $T_2^\perp$ are possible within the initial demagnetization [see $p_{C/A}$, gray dashed line in Figs.~\ref{fig:relative_rates_initial_steps}(e,f)]. 
%
The formation of coral-shaped domains can thus be directly linked to the emergence of $T_2^\perp$ from "charge separation" events within the propagation of $T_1$ strings via transitions of type C.\cite{2013Farhan} As discussed above, transitions of type C are promoted within the chiral-split barrier model, without need to invoke site-specific disorder.
Therefore, the experimental observation of coral-shaped domains indicates the effect of chiral barrier splitting, in addition to the influence of lithographic defects to which these morphologies are typically attributed.\cite{2013Farhan,BUDRIKIS2014,GILBERT2015,JENSEN2022}

\subsection{Relaxation at Domain Boundaries}
\label{sec:transitions_at_domain_boundaries}

As shown in Fig.~\ref{fig:demagnetisation}, on a macroscopic scale, relaxation slows considerably for $M_{[11]}\le$25\%, indicating that the relevant switching environments and relaxation pathways effectively change, now being largely constrained to domain boundaries only. At the same time, especially in the case of the chiral-split barrier model or weak interactions, a sizeable number of $T_3$ vertices arise with a peak around $M_{[11]}\approx$10\%, emphasizing the vital role of high-energy vertices in the final relaxation steps to the ground state. 

The cyan and blue shaded areas in Fig.~\ref{fig:demagnetisation}(a-d) representing the population of $T_2^\perp$ and $T_2^{\overline{1}\overline{1}}$ vertices (magnetized opposite to the initial magnetization) furthermore indicate a specific sequence for the final relaxation to the ground state, by the creation of extended $T_1$ domains and their removal via mobile $T_3$ vertices (red area): 
The emergence of $T_2^\perp$ vertices not only relates to the emergence of proto-domains but also predates the onset of a sizable population of $T_2^{\overline{1}\overline{1}}$ vertices. 
The onset of a $T_2^{\overline{1}\overline{1}}$ vertex population also coincides roughly with the maximum of the $T_3$ population (red area), as their motion and annihilation are crucial for the consolidation of the ground-state domains. 
Together with the original $T_2^{[11]}$ vertices, a large fraction of these emerging $T_2^\perp$ and $T_2^{\overline{1}\overline{1}}$ vertices compound domain boundaries.

We now consider a prototypical coral domain boundary as shown in Fig.~\ref{fig:domains}(a). 
This boundary consists mostly of $T_2$ vertices as well as $T_3$ pairs,\cite{MORGAN2011a,BUDRIKIS2012} with a straight or zigzag magnetization following the domain outline (indicated by white lines).\cite{JENSEN2022}
Different transitions, depending on the local spin configuration, are marked with letters in Fig.~\ref{fig:domains} and moments for which $M^\perp\ne0$, i.e., with transitions boosted by barrier splitting in the CB model, are marked with black arrows [see Tab.~\ref{tab:transitions} for values of $\Delta E^\text{dip}_{i\rightarrow f}$ and $M_i^\perp$].

Excitations of the ground state, i.e., $T_1T_1\rightarrow T_3T_3$ (transition K in Tab.~\ref{tab:transitions}), are energetically costly. Therefore, within the last stages of relaxation, the switching events are largely localized to the domain boundaries, further contributing to the slowdown of dynamics in the final stage of relaxation (as shown in Fig.~\ref{fig:demagnetisation}).
%
Many potential spin flips at the boundary furthermore involve a large barrier energy (labeled in light gray in Fig.~\ref{fig:domains}), especially in the case for diagonal domain boundaries, as many of these promote the creation or motion of $T_3$ vertices. 
%
Even when considering chiral effects, many transitions remain energetically unfavorable (black arrows denote those environments that experience torque exerted by neighboring spins). 

For \textit{straight} boundaries, Fig.~\ref{fig:domains}(b), the creation of a $T_3$ pair leads to an asymmetry with respect to the two different ground state domains on either side of the boundary. Transition L would create a $T_4$ vertex and thus is kinetically suppressed, whereas transitions J and H with $\Delta E_{i\rightarrow f}^\text{dip}=0$ support a high mobility of the $T_3$ defects to move within the boundary line or deform the domain.

For \textit{diagonal} domain boundaries, Fig.~\ref{fig:domains}(c), the presence of $T_3$ vertices inevitably comes hand in hand with domain deformation. Similarly to the straight boundary, a clear asymmetry in transitions contributes to the preferential shrinking of one ground-state domain over the other. Note that A* transitions are unlikely to occur within a mean-field model but can be promoted via chiral-barrier splitting in the limit of strong interactions (i.e., small lattice periodicities $a$).

Within our kMC simulations, we rarely reached full ground-state ordering within $n^3$ simulation steps due to a high fraction of random transitions (for weaker interactions) or limiting relaxation pathways. The latter is especially the case for the mean-field model with strong interactions, where size-dependent periodic boundary conditions may further reduce chances of full demagnetization. 
The many spin flips necessary to reach the ground state might explain the prevalence of coral-shaped domains observed in experiments where spin configurations are frozen after field-driven or thermal annealing.

\section{Discussion}

%
In this work, we demonstrated how the overall relaxation towards the energetically favorable ground state can be modified profoundly by explicitly taking modifications from local torques to kinetic energy barriers into account. 
Although the main relaxation (i.e., reduction of the net magnetization from 100\% to 25\%) involves a small number of relevant transitions only, a subtle shift in their relative probability results in different emergent mesoscopic features, ranging from string-like avalanches to coral-shaped domains in a perfect system. 
%
%
As our model considers intrinsic effects only in an otherwise idealized homogeneous s-ASI, the emergence of diverse domain structures implies a complementary viewpoint on effects typically attributed to extrinsic disorder by assuming a site-specific variation in the value of $\Delta E_\text{sb}$, often around 5\%\cite{2013Farhan,BUDRIKIS2014,GILBERT2015,JENSEN2022}.


Common experimental techniques used to measure and image artificial spin systems do not have the temporal and spatial resolution to directly track the preferred rotation direction of switching nanomagnets. 
Hence, we have to consider indirect measures on how the chiral-split barrier model will produce better predictions for experimental systems when compared to the mean-field model, especially when considering additional randomizing effects due to extrinsic disorder. 

We briefly discuss several experimental observations that could indicate an influence of kinetics affected by spin flips with favorable switching chirality. 

First, the ratios between the transition rates A, B and C determine the average length of $T_1$ strings, as well as the initial rate of demagnetization; however, these changes are subtle and difficult to disentangle from the effects of extrinsic disorder or temperature. 

Second, the increased back-flip probability for $T_3$ pairs could show up in experiments as an unusual persistence of the ferromagnetic state.\cite{BINGHAM2022}

Third, experimental results often indicate an arrested state around 20\% of the final magnetization,\cite{2013Farhan,PORRO2013,MORLEY2017} at which the relaxation dynamics slow down rapidly. Inspecting domain morphologies and vertex populations (in particular $T_3$, $T_2^\perp$ and $T_2^{\overline{1}\overline{1}}$), as well as site-specific blocking temperatures,\cite{2010Li,GILBERT2016, 2020Keswani} could indicate the relative importance of the different relaxation pathways leading to these states (without ascertaining whether they originate from disorder or chiral-split barriers).

Fourth, and finally, quantifying the switching activity at boundaries between ground-state domains\cite{MORGAN2011a,BUDRIKIS2012,2019Chen_a,JENSEN2022} might give crucial insights to favored transitions and how they relate to local disorder compared to the effect of barrier reductions due to local torques.

\section{Conclusions}

In conclusion, in this work we consider the effect of different models based on a point-dipole description of switching barriers in s-ASI, and follow the evolution of a perfect system using kinetic Monte Carlo simulations. 
We find relaxation dominated by avalanches of $T_1$ strings in the limit of the (static energy-only) mean-field barrier model, contrasted with the emergence of more complex coral-shaped domains by taking into account the effect of local torques that lead to favored chiral switching pathways.
We find differences between the models in particular with respect to their influence on initial demagnetization, the role of $T_2^\perp$ vertices in the formation of initial domains, and relaxation pathways at domain boundaries.

Furthermore, we find that analyzing single vertex populations, typically $T_3$ in literature, is not as informative as considering also the sizable fraction $T_2^\perp$ and $T_2^{\overline{1}\overline{1}}$ vertices. These vertex distributions develop concomitantly, and thus give insight into the relative importance of different relaxation pathways.

Our results make a step towards creating more realistic, but still highly simplified, models to describe the switching in experimental artificial spin systems dominated by coherent reversal modes and strong pairwise interactions. 
Although we here focus on the s-ASI with periodic boundary conditions, where each moment has the same environment, the effect of the local torques would be even more pronounced in the case of artificial spin systems with mixed-vertex coordination and different boundary conditions.\cite{MORRISON2013,GILBERT2014,GILBERT2016,LAO2018,SAGLAM2022}

Descriptive models that capture the relevant features of transition kinetics are key for successful simulations of the evolution of large arrays, which are otherwise too costly or plainly unrealistic to simulate with micromagnetic approaches. 
Simulation tools based on simplified but physically realistic descriptions\cite{JENSEN2022,2024Maes} therefore make it possible to successfully model the evolution of extended spin ices\cite{2024Jensen} or develop sample geometries tailored to specific nanomagnetic logic or unconventional computing applications.\cite{2018Arava,ARAVA2019,ALLWOOD2023,2024Stamps}

\section*{Data Availability}
The data set supporting the data presented in this work can be found in the zenodo archive with DOI \href{https://doi.org/10.5281/zenodo.15548088}{10.5281/zenodo.15548088}.

\begin{acknowledgements}
	M.M.\ and P.V.\ acknowledge support from the Spanish Ministry of Science, Innovation and Universities under the Maria de Maeztu Units of Excellence Programme (CEX2020-001038-M) and the pre-doctoral Grant PRE2019-088070 (M.M.), as well as from the Spanish Ministry of Science, Innovation and Universities and the European Union under the project number PID2021-123943NB-I00 (OPTOMETAMAG).
	
	N.L.\ received funding from the European Research Council (ERC) under European Union's Horizon 2020 research and innovation program under the Marie Sklodowska Curie Grant Agreement No.~844304 (LICO\-NAMCO), as well as support by a UKRI Future Leader Fellowship MR/X033910/1 (LIONESS).
\end{acknowledgements}

%




\clearpage
\newpage
\beginsupplement
\title{\papertitle\linebreak Supplemental Material}
\maketitle

\begin{figure}
	\includegraphics{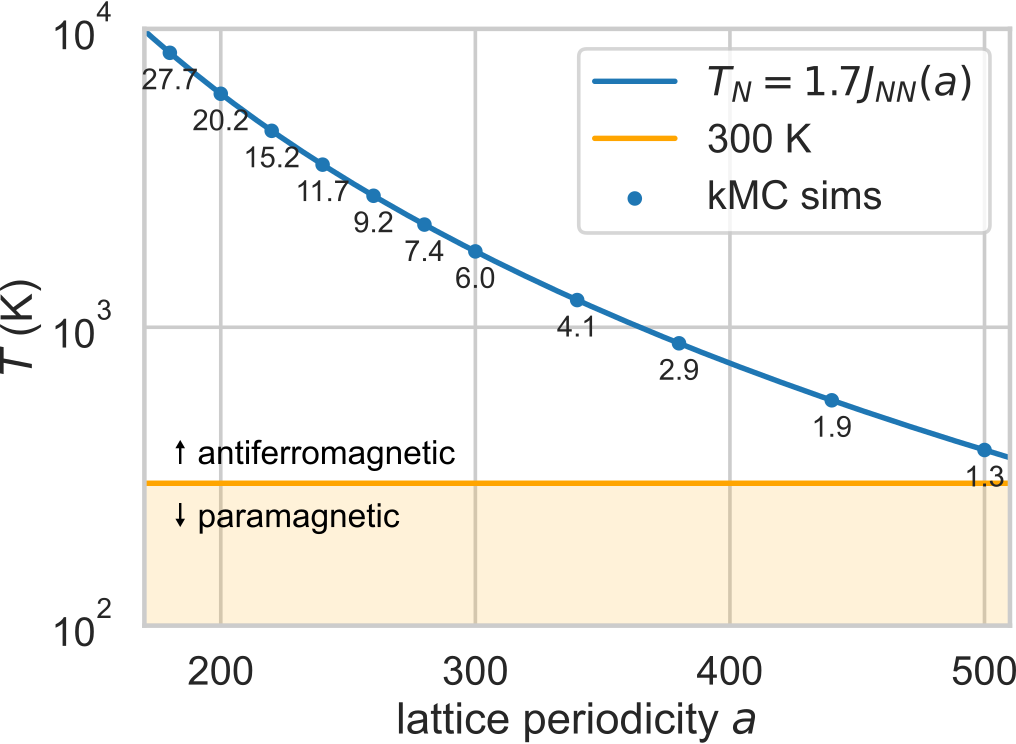}
	\caption{%
		\textbf{Néel temperature in dependence of lattice periodicity $a$}. 
		The blue curve denotes the critical temperature $\TN(a)=1.7\JNN$ in dependence of lattice periodicity $a$, calculated using Eq.~(\ref{eq:JNN}) for $m=\num{3.285e6}\mu_\mathrm{B}$. Points denotes values of $a$ for which kMC simulations were performed, the numbers underneath each point denote the ratio $\TN/T$, i.e., how far within the ordered phase the demagnetisation was simulated.
		}
	\label{sfig:TN-vs-a}
\end{figure}

\begin{figure*}
	\includegraphics[width=\textwidth]{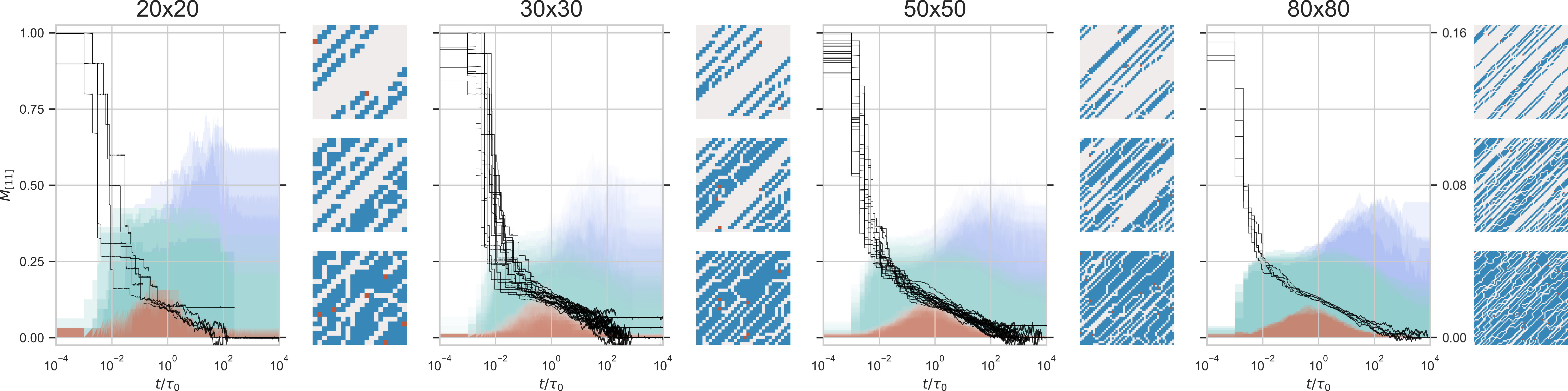}
	\caption{%
		\textbf{kMC simulations for the split-barrier model in dependence of system size,} for fixed $a=\SI{260}{nm}$ and $T=\SI{300}{K}$, and $n\times n$ vertices with $n=(20,30,50,80)$ comparing the results of $(5,20,20,5)$ kMC simulation runs, respectively. All other simulations settings are as described in the main manuscript.
		Equivalent to Fig.~\ref{fig:demagnetisation}, the four panels compare the time-dependence of demagnetisation (black lines), $T_3$ vertex population (red), as well as $T_2^\perp$ (cyan, from zero) and $T_2^{[\bar{1}\bar{1}]}$ (blue). Insets on the right show representative spatial configurations of vertices at times where the net magnetization is at 75\%, 50\% and 25\% of its initial value. 
		Finite-size effects are clearly visible in the case of $n=20$, whereas the results of $n\ge30$ resemble each other well. Balancing better statistics with computational resources, we chose to discuss the kMC results of $n=50$ for the main manuscript, and used a system size of $n=30$ for the phase diagram shown in Fig.~\ref{sfig:phase_diagram}.
		}
	\label{sfig:system_size_comparison}
\end{figure*}

\begin{figure*}
	\includegraphics{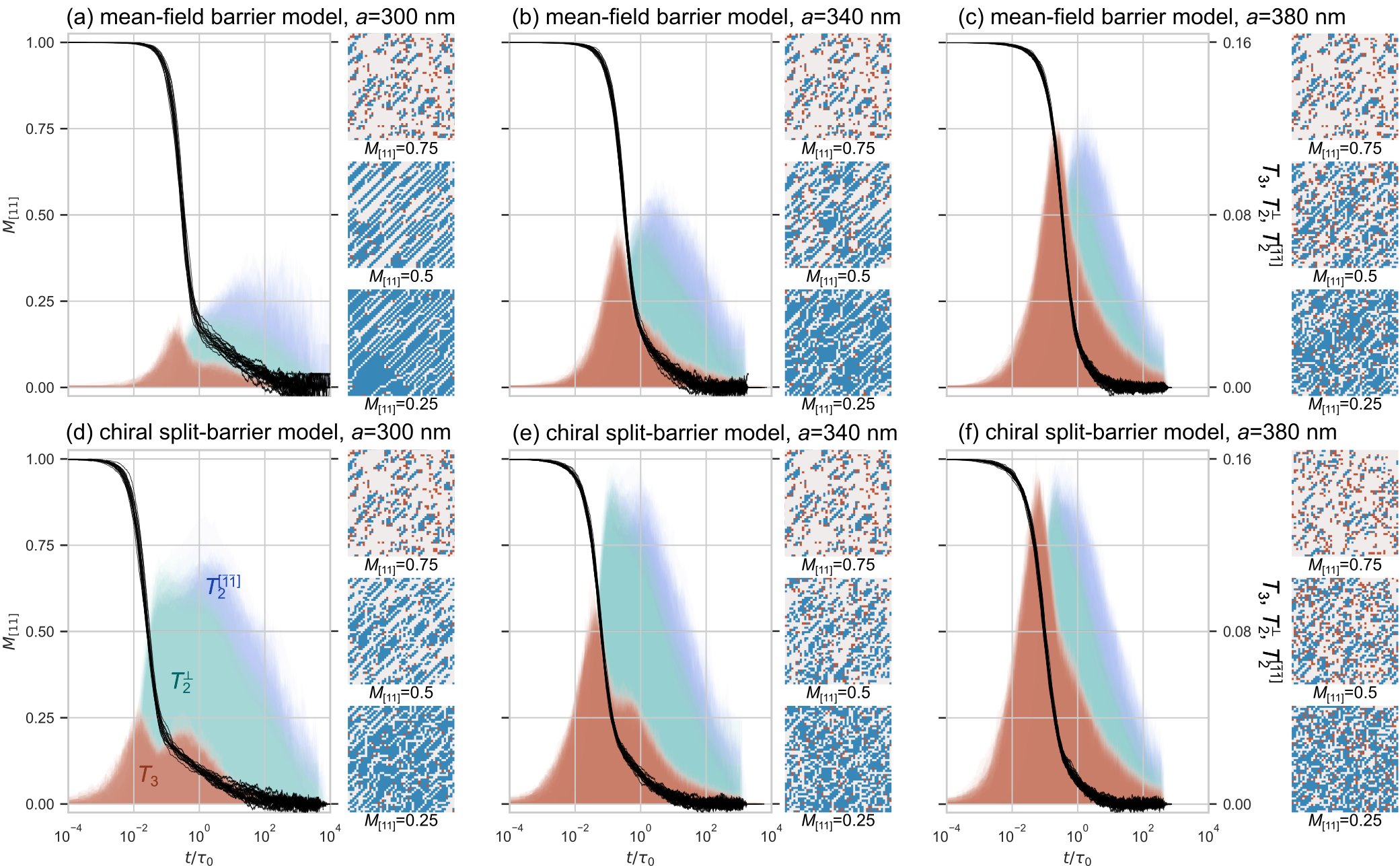}
	\caption{%
		\textbf{Demagnetization behaviour for lattice periodicities $a$ above \SI{300}{nm}} for (a-c)~the mean-field barrier model and \mbox{(d-f)}~the chiral-split barrier model. Legend and annotations are equivalent to Fig.~\ref{fig:demagnetisation} of the main manuscript. 
	}
	\label{sfig:demagnetisation-large_a}
\end{figure*}

\begin{figure*}
	\includegraphics{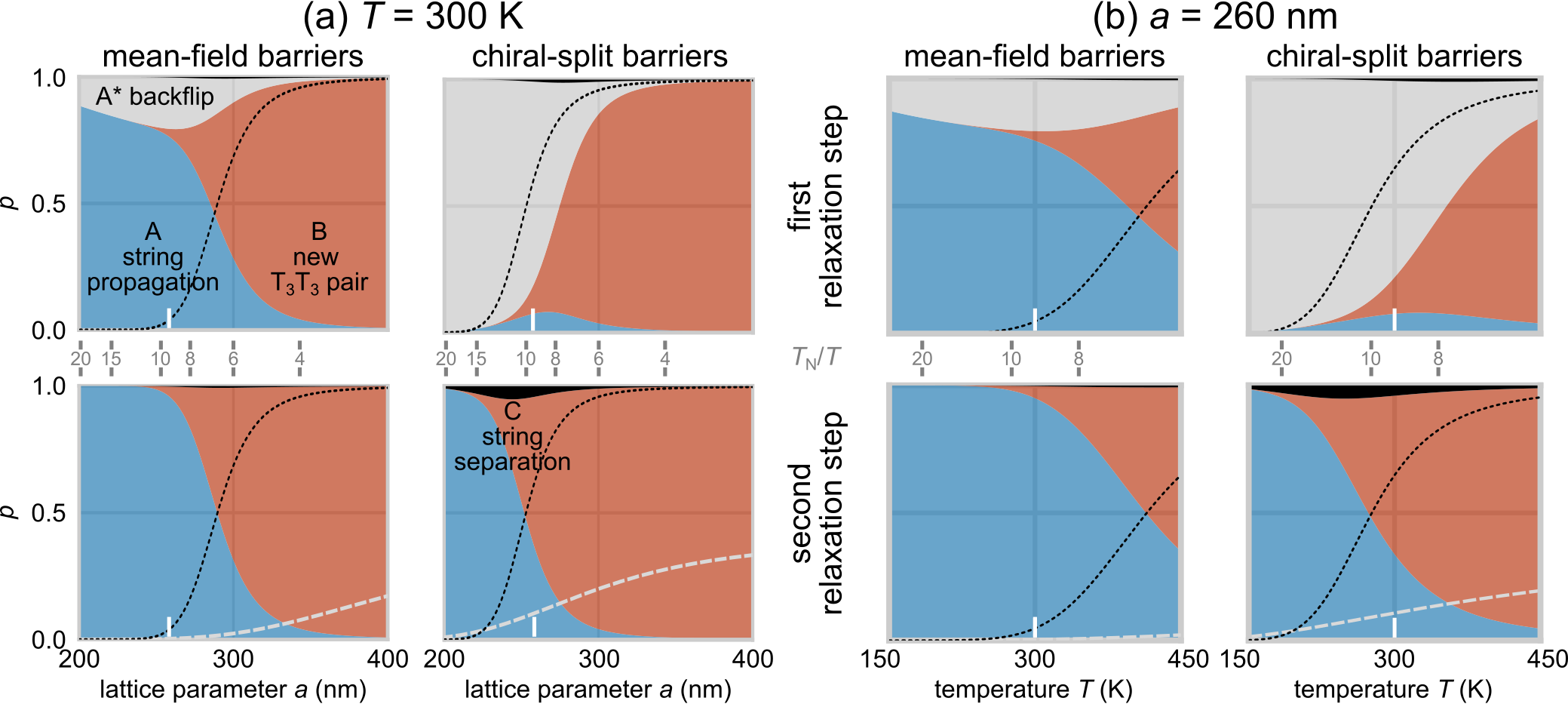}
	\caption{%
		\textbf{Equivalence of periodicity and temperature for the initial demagnetisation steps.}
		Relative probabilities of the four most likely transitions upon the first (top row) and second (bottom row) relaxation steps, for either (a) constant temperature $T=\SI{300}{K}$ and (b) constant periodicity $a=\SI{260}{nm}$ (white markers), comparing the mean-field barrier model (left) to the chiral-split barrier model (right). 
		%
		%
		The gray numbers in the middle denote the ratio $T_\mathrm{N}/T$ between the Néel temperature $\TN$ and $T=\SI{300}{K}$. Comparing the curves at similar ratios, the variation with $a$ at fixed $T$ in panel~(a) is equivalent to the curves for varying $T$ at fixed $a$ in panel~(b).
		}
	\label{sfig:initsteps_a_vs_T} 
\end{figure*}

\begin{figure*}
	\includegraphics{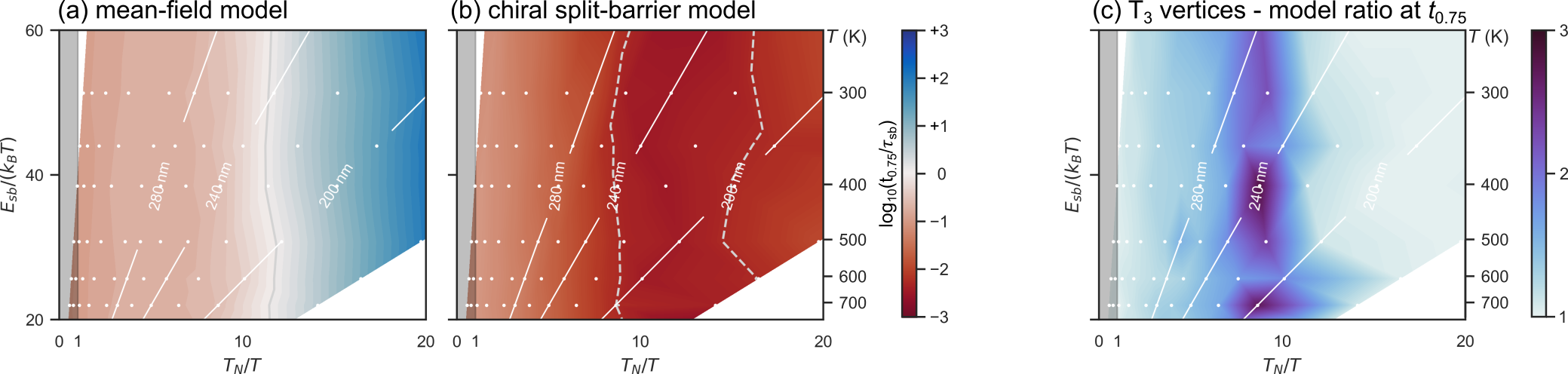}
	\caption{%
		\textbf{Generalised phase diagram,} calculated from kMC simulations at different lattice constants $a$ (\SIrange{170}{500}{nm}) and temperatures $T$ (\SIrange{200}{800}{K}) at a system size of $n=30$ (20 averages each). All other parameters, i.e., $m$ and $\Delta E_\mathrm{sb}$, remain as described in the main manuscript.
		%
		To allow for a more general comparison, values are plotted for each simulation (white points) in dependence of the relative single-particle barrier $\Delta E_\mathrm{sb}/(\kB T)$ [corresponding values of $T$ are shown on the right scale of (b)] and relative interaction temperature $\TN/T$ [values of constant $a=$\SI{200}{nm}, \SI{240}{nm} and \SI{280}{nm} are marked with white lines, the paramagnetic region is shaded in gray]. 
		\newline
		(a,b)~Demagnetisation time $t_{0.75}=t(M_{[11]}=75\%)$ divided by $\tau_\mathrm{sb}$ for the (a)~mean-field and (b)~split-barrier model [comparable to Fig.~\ref{fig:quantile_times}]. 
		(a)~For the mean-field model and $\TN/T\gtrapprox 11$ the relaxation shows a slow down (gray line, blue region) even beyond the speed of single-particle relaxation due to correlated string propagation. (b)~For the split-barrier model and $\TN/T\gtrapprox 16$ the initial demagnetisation is delayed due to rapid creation-annihilation events of $T_3$ pairs (dashed lines).
		\newline
		(c)~Ratio of the number of $T_3$ vertices at $t_{0.75}$ between the split-barrier  and mean-field barrier model. For values $\TN/T\approx 9$, two to three times more $T_3$ vertices are created for the split-barrier, coinciding with an enhancement of string separation events (as shown in Fig.~\ref{sfig:initsteps_a_vs_T}), leading to complex domain structures.
		}
	\label{sfig:phase_diagram}
\end{figure*}

\end{document}